\RequirePackage{amsmath} 
\documentclass[runningheads]{llncs}
\pdfoutput=1
\usepackage{graphicx}
\usepackage{color}

\usepackage[squaren,Gray]{SIunits}
\usepackage{array}
\usepackage{multirow}
\newcolumntype{L}[1]{>{\raggedright\arraybackslash }m{#1\textwidth}}
\newcolumntype{C}[1]{>{\centering\arraybackslash }m{#1\textwidth}}

\begin{document}
\title{DistNet: Deep Tracking by displacement regression: application to bacteria growing in the {\itshape Mother Machine}}
\titlerunning{DistNet: Deep Tracking by displacement regression}
%

\author{Jean Ollion\inst{1}\inst{2}\orcidID{0000-0002-9182-4624} \and
Charles Ollion\inst{3}\orcidID{0000-0002-6763-701X}}
\authorrunning{J. Ollion and C. Ollion}
%
\institute{
Laboratoire Jean-Perrin UMR 8237, Sorbonne Universit\'e, Paris, France 
\and 
SABILAb, Paris, France \\
\url{https://sabilab.github.io/} \\
\email{jean.ollion@polytechnique.org} 
\and
CMAP, Ecole Polytechnique, Universit\'e Paris-Saclay \\ \email{charles.ollion@gmail.com}}
\maketitle              
\begin{abstract}

The {\itshape mother machine} is a popular microfluidic device that allows long-term time-lapse imaging of thousands of cells in parallel by microscopy. It has become a valuable tool for single-cell level quantitative analysis and characterization of many cellular processes such as gene expression and regulation, mutagenesis or response to antibiotics.
The automated and quantitative analysis of the massive amount of data generated by such experiments is now the limiting step. 
In particular the segmentation and tracking of bacteria cells imaged in phase-contrast microscopy---with error rates compatible with high-throughput data---is a challenging problem.

In this work, we describe a novel formulation of the multi-object tracking problem, in which tracking is performed by a regression of the bacteria's displacement, allowing simultaneous tracking of multiple bacteria, despite their growth and division over time. 
Our method performs jointly segmentation and tracking, leveraging sequential information to increase segmentation accuracy.

We introduce a Deep Neural Network (DNN) architecture taking advantage of a self-attention mechanism which yields extremely low tracking error rate and segmentation error rate. We demonstrate superior performance and speed compared to state-of-the-art methods. Our method is named DiSTNet which stands for DISTance+DISplacement Segmentation and Tracking Network.

While this method is particularly well suited for {\itshape mother machine} microscopy data, its general joint tracking and segmentation formulation could be applied to many other problems with different geometries.
\end{abstract}

\section{Introduction}

\subsection{Context}
\begin{figure}[!ht]
    \centering
    \def\svgwidth{0.9\textwidth}
    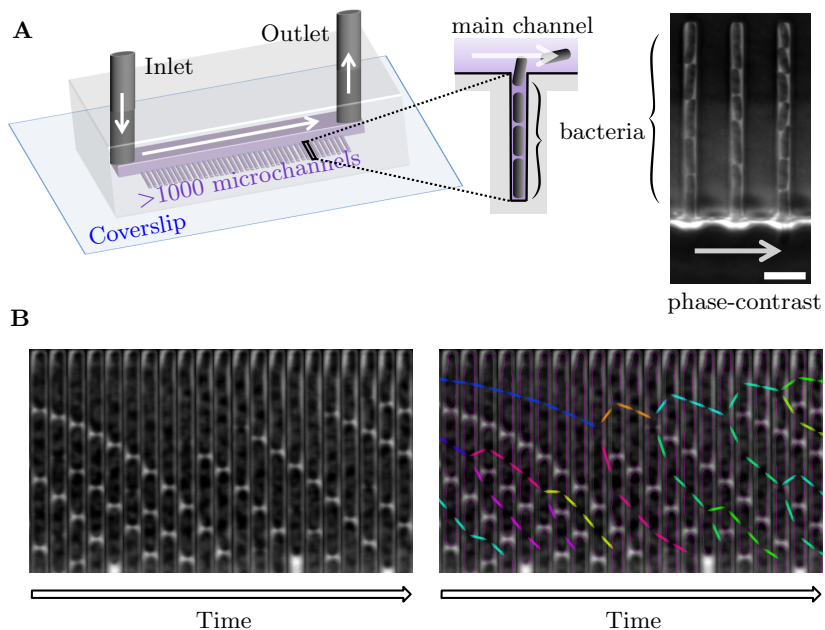
    \caption{{\itshape Escherichia coli} bacteria growing in the {\itshape mother machine}. A: left pane: the {\itshape mother machine} microfluidic chip; right pane: corresponding phase-contract microscopy images; scale bar: \unit{5}{\micro\meter}; white arrows represent the flow of growing medium. B: Kymographs showing phase-contrast images of bacteria growing in a microchannel; images from successive frames are displayed next to one another; cells are going out through the open ends located at the lower part of the images. The right pane displays outlines of segmented bacteria and tracking links as coloured arrows, each colour representing one generation.}
    \label{figkymo}
\end{figure}

Single-cell study has become a focus of research in numerous fields of biology during the past decades \cite{davie2018single,elowitz2002stochastic,navin2011tumour}.
In particular, time-lapse microscopy has been extensively used to investigate cellular processes dynamically and non-invasively in single cells \cite{muzzey2009quantitative}.
It is now being increasingly used in combination with microfluidic devices that allow both high-throughput data collection at the single-cell level and a precise spatiotemporal control of the environment \cite{mehling2014microfluidic}.
Among those devices, the {\itshape mother machine}, developed by Wang et al. in 2010 \cite{wang2010robust}, is one of the most popular. It contains thousands of parallel dead-end microchannels in which cells grow in single file (See Fig.~\ref{figkymo}). Cells can grow and divide inside the microchannels for hundreds of generations, allowing the imaging of $10^5 - 10^6$ individual cells per experiment. {\itshape Mother machine} devices are being increasingly used for single-cell studies on bacteria to investigate various subjects, such as gene expression and regulation \cite{norman2013memory,brenner2015single,kaiser2018monitoring}, mutagenesis and evolution \cite{robert2018mutation,robert2019real} or single cell response to antibiotics \cite{bamford2017investigating,bergmiller2017biased}. 
The massive amount of data generated by long-term imaging of cells growing in the {\itshape mother machine} (typically several hundred gigabytes worth of images per experiment) needs to be processed automatically, in particular methods for automatic segmentation and tracking of cells with very low error rate are needed. 

\subsection{Problematic}
In this work, we focus on segmentation and tracking of bacteria growing in microchannels observed with phase-contrast, a very common imaging technique.
The term tracking refers to the matching of the observed bacteria between two successive frames, as shown with colored arrows in Fig.~\ref{figkymo}-B.

Tracking of bacteria growing in the {\itshape mother machine} faces three major challenges: 
\begin{itemize}
    \item cell growth induces changes in bacteria morphology;
    \item bacteria can divide;
    \item due to cell growth, bacteria located at the open-end of microchannels are pushed out by other bacteria, thus their next observation is sometimes outside or partly outside the image.
\end{itemize}

Studying some biological processes such as mutagenesis require very fine statistics in order to detect rare events as in \cite{robert2018mutation}. To achieve this, one need to analyse massive datasets with typically $10^6 - 10^7$ observations of bacteria, at a very low error rate, typically less than $0.01\%$, in order to limit manual curation time.


\subsection{Related work}

Multi-object tracking is a challenging task well studied within the computer vision community. Multi-object detection and tracking are usually considered as separated processes \cite{luo2014multiple,chen2019mot}, where object detection on natural images have much improved, see e.g. Faster-RCNN \cite{ren2015faster}. Tracking then relies on finding similarities between detections ---usually a combination of the semantics (deep features), shape and velocity of the detected objects--- and matching them over the successive frames~\cite{zhang2008global,yu2016poi}. 
A few studies perform both tasks in a single pass, such as TrackNet \cite{li2019tracknet}, which derives from the Faster-RCNN architecture. While the method yields promising results, it remains very complex and the authors admit that the simultaneous detection and tracking problem is still at its infancy. 
Moreover, these methods do not take into account specific aspects of context of bacteria growing in microchannels, such as the bacteria division.

Several methods have been developed specifically for this problem, most of them perform segmentation and/or tracking with a combination of pre-defined classical computer vision operations \cite{kaiser2018monitoring,sachs2016image,ollion2019high,smith2019mmhelper} and are thus very difficult to tune for datasets generated on different imaging setups and/or strains. 
A recent software, DeLTA \cite{lugagne2019delta}, uses DNN both for segmentation and tracking. Segmentation is performed first, using the original U-Net approach \cite{ronneberger2015u} the tracking is performed by another U-Net-like neural network, which predicts the next cell(s) for each cell.
The authors report an error rate of $1\%$, which can be too high for some applications in which rare events are studied for instance.

\section{Method}

\subsection{Problem formulation}

The main contribution of this work is to track bacteria by performing a regression of their displacement between two successive frames. This formulation contrasts with previous tracking systems, as the detection and tracking are performed simultaneously and in one pass for all bacteria. This is motivated by the following:
\begin{itemize}
\item {\bfseries Global consistency:} we expect this method to have more coherent results, i.e. less conflicting predictions compared to a method that make one prediction per bacterium, because tracking is done simultaneously for all bacteria. 
\item {\bfseries Speed:} This method is faster because one prediction by image is needed instead of one for the detection, then a second one per bacterium for the tracking.
\item {\bfseries Simplicity:} Our method enables to jointly train a single model, which is derived straightforwardly from a U-Net architecture, and could be adapted easily to different problem settings or backbone networks. In contrast, tracking methods involving two steps and several models induce more hyper-parameters and complexity. Note that adapting to models such as Faster R-CNN (not covered in this study) would require further developments. 
\end{itemize}

\subsubsection{Model description}
As bacteria grow in a single file in the microchannels, prediction of the displacement along the axis of microchannel (further called Y-axis) is sufficient.
We achieve this by predicting, for each bacteria observed frame $F$ the displacement along the Y-axis between its center and the center of the same bacteria observed at frame $F-1$ (See Fig.~\ref{figunet}-(i)).
Formally, we predict a map $m_{x,y}$ which has the same spatial dimensions as the frame $F$, defined as follows:
\begin{equation}
m_{x,y}=
\begin{cases}
  c^F(B) - c^{F-1}(B), & \text{if} \ (x,y) \  \text{inside bacterium} \ B \\
  0,                   & \text{if} \  (x,y) \  \text{otherwise}
\end{cases}
\end{equation}
where $c^F(B)$ is the Y-coordinate of the center of bacterium $B$ at frame $F$.

Tracking at a given frame $F$ is then simply achieved by moving each bacteria observed at a frame $F$ by the opposite value of their predicted displacement and associating them to the most overlapping bacterium at frame $F-1$. 
We make a prediction of the previous observation of the bacteria and not of the next, because bacteria can divide but not merge and thus each bacteria can be associated to at most one single bacteria observed at the previous frame.

Another important aspect of our method is that we perform segmentation jointly with tracking by a regression of the euclidean distance map (EDM), Fig.~\ref{figunet}-(c-d). Segmentation by regression of the EDM has been proposed before \cite{graham2018hover,naylor2018segmentation,heinrich2018synaptic}: in contrast to the original U-Net formulation that focuses on cell contours, it detects the interior of the cells. It is likely to be more robust in cases where cell contours are less visible, which happens very often in the case of bacteria growing in microchannels when they are in close contact to one another or to the border of the microchannel (See Fig.~\ref{figkymo}). A previous study suggests that this formulation pushes the neural network to learn some notion of objects \cite{naylor2018segmentation}, which may also benefit to tracking.
Individual cells are then easily segmented using a classical watershed transform as in \cite{naylor2018segmentation}. As EDM corresponds to the distance to background (in pixels unit), the watershed is naturally restricted to predicted EDM$\geq1$. To limit over-segmentation, segmented regions in contact with each other where merged when the EDM value at their interface was over a threshold.

The network also predicts a category: background, dividing cells, cell not linked to a previous cell, other cells (non dividing and linked to a cell at previous frame), as exemplified in Fig.~\ref{figunet}-(e-h). When a cell is predicted as having no previous cell---such as the bottom cell in Fig.~\ref{figunet}-(d)---it is not linked to any cell in the tracking procedure.

\subsection{Network architecture}
\begin{figure}[!ht]
    \centering
    \def\svgwidth{0.9\textwidth}
    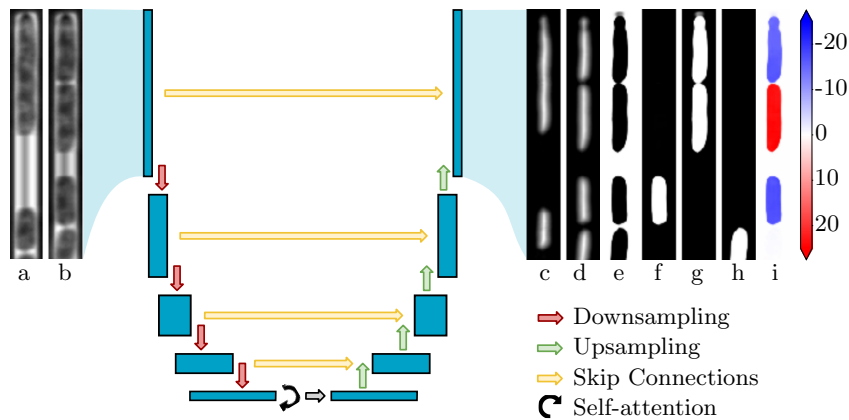
    \caption{U-Net architecture. Each blue block corresponds to a 2D multi-channel feature map. The network has an encoder-decoder structure. The encoder reduces spatial dimensions and increases the number of channels at each contraction (red arrows). The decoder reduces the number of channels at each up-sampling level, and restores spatial dimensions using both feature maps of the previous level (green arrows) and of the corresponding level in the encoder (yellow arrows). Our modified version of U-Net contains a self-attention layer at the last contraction level (see section~\ref{sisa} for mathematical details). Inputs are couples of successive grayscale images (a: previous frame, b: current frame). The upper bacteria in (a) divides in (b). Outputs are: EDM predictions for the previous (c) and current frame (d); category prediction (e): background, (f): cell that do not divide and are associated to a cell at the previous frame, (g): cells that divided, (h): cells that are not associated to a cell at the previous frame); (i): prediction of bacteria Y-displacement between the two frames, in pixels and within an image of height 256.}
    \label{figunet}
\end{figure}
While U-Net-like architectures are efficient to integrate local semantic information to take precise decision at the pixel level \cite{ronneberger2015u} (See Fig.~\ref{figunet}), we posit that they lack understanding of global structure. This is not particularly a problem for semantic segmentation as a rather restricted context might be enough to provide the necessary object boundaries information. However, in the context of tracking objects with potential division or displacement, some decisions must be made at the global image level, so that no contradictory information emerge from two different local contexts - for instance a very large bacteria dividing into two should have coherent global behavior, rather than independent prediction for each of the local bacteria.

In order to integrate global context, and following recent success in natural language processing, graph-based machine learning as well as many other fields, we incorporated a self-attention layer similar to the ones found in the {\itshape Transformer} architecture \cite{vaswani2017attention} instead of the the last convolution of the encoder network. This layer enables the DNN to combine information from the whole image, while a convolution only mixes information locally. The description of the layer is detailed in section~\ref{sisa}.

As a reference, we compared this strategy to a previously proposed one that consists in stacking hourglass models (i.e. encoder-decoder architectures) \cite{newell2016stacked}. This model is referred to as stacked hourglass model and described in section~\ref{sistacked}.

\section{Experiments}

\subsection{Training}

We created a training dataset (referred to as DST) of 65344 images from 3 different {\itshape E. Coli} strains acquired on the same setup (from \cite{robert2018mutation}). We used BACMMAN software \cite{ollion2019high} to generate labels and curate them manually.
In order to limit over-fitting and increase generalization to different domains (experimental setups, strains or mutants), we performed a data augmentation step with both classical and specifically designed transformations described in section~\ref{sidataaug}. 
Model architectures are detailed in section~\ref{sitraining}.

\subsection{Evaluation}
We created two evaluation datasets composed of images of randomly chosen microchannels from several experiments performed with different imaging setups and/or strands using the same procedure as for DST. The first one (referred to as DSE1) corresponds to experiments where bacteria have similar aspect to those of DST (8 different experiments including 3 different strands and 2 different setups, 51000 observations of bacteria within 12600 frames). Images of DST and DSE1 were chosen among distinct microchannels. The second one (referred to as DSE2) to experiments where bacteria have an aspect that differs substantially from DST (3 experiments including 3 different setups and strands, 8760 observations of bacteria within 1680 frames). DSE2 is composed of images from publicly available datasets published along with software for {\itshape mother machine} analysis \cite{sachs2016image,lugagne2019delta,kaiser2018monitoring}. We excluded frames with important anomalies on bacteria morphology. 

We analysed 3 types of segmentation errors: false positive (predicted bacteria with no corresponding ground truth bacteria), false negative (ground truth bacteria with no corresponding prediction) and division errors (DE). The latter occur when a division is detected too early or too late. As the exact frame at which a cell divides is sometimes hard to discern visually, we added a tolerance of one frame, i.e. we counted an error if a division was detected at least two frames before or two frames after the ground truth division frame. Matching between ground truth cells and predicted cells was done using a maximum overlap criterion, and insignificant effect is observed when using a IoU-based criterion instead.
We did not use metrics that estimate the overlap with ground truth such as IoU because exact cell contours are difficult to obtain, which makes these metrics less informative.
We also analysed tracking errors, which occur when a link is predicted between two bacteria and that the corresponding bacteria in the ground truth are not linked. Note that the links that differ between ground truth and prediction as a consequence of a DE are not considered as tracking errors as long as linked bacteria remain in the same lineage.
We excluded bacteria that were partially out of the images (i.e. going out of the microchannels) with a length lower than 40 pixels, because they are often excluded from analysis, in an automated way.

\section{Results}
We compared the performance of our method to two baselines: segmentation and tracking performed by BACMMAN and DeLTA softwares. DeLTA models were also trained on the same dataset as our model with the same data augmentation scope. Table~\ref{tableevalsim} shows the percentage of errors on dataset DSE1: both self-attention and stacked hourglass models perform better than the two baselines for segmentation and tracking. Importantly both models achieve very low error rates, lower than $0.005\%$ for tracking and of $0.03\%$ for segmentation. The Self-attention network displays slightly better performances than the stacked hourglass network in terms of accuracy and speed, and has less parameters. 
The percentages of segmentation and tracking errors of DeLTA on DSE1 are compatible with those reported by the authors \cite{lugagne2019delta}. The higher number of tracking errors made by DeLTA is mainly due to contradictory predictions that happen when the same next cell is predicted for two different cells.

\begin{table}
    \caption{Percentage of errors on dataset DSE1 (cells with similar aspect to the training set). Last column shows inference time in seconds for 1000 frames, on CPU (2 Intel Core i7-4790K 4GHz) / GPU (GeForce RTX 2080 Ti).}
    \begin{tabular}{L{0.22}C{0.13}C{0.13}C{0.1}C{0.1}C{0.11}|C{0.16}}
        \hline
        \multirow{2}{*}{\bfseries Method/Model}&
        {\bfseries Tracking}&
        \multicolumn{3}{c}{\bfseries Segmentation Errors}&
        \multirow{2}{*}{\bfseries Total}&
        {\bfseries Execution}\\\cline{3-5}
        &{\bfseries Links}&{\bfseries Division}&{\bfseries False $-$}&{\bfseries False $+$}&&{\bfseries Time}\\
        \hline
        BACMMAN & $0.14$ & $0.63$ & $0.036$ & $0.0076$ & $0.82$ & $23/NA$\\ 
        DeLTA & $0.57$ & $0.069$ & $0.0018$ & $0.069$ & $0.71$ & $249/6$\\ 
        Stacked Hourglass & $0.0021$ & $0.025$ & $0.0038$ & $0$  & $0.031$ & $126/5$\\
        Self-Attention & $0.0042$ & $0.019$ & $0.0019$ & $0$  & $0.025$ & $100/3$\\
        [0.5ex]
        \hline\\
    \end{tabular}
    \label{tableevalsim}
\end{table}

In order to estimate the generalization capacity of our network, we performed an evaluation on dataset DSE2, which contains bacteria displaying a different aspect from DST. Percentage of errors are shown in Table~\ref{tableevaldiff}. We observe a significant drop of performances for all methods, to $7-8\%$ for the baseline methods and $1\%$ for our method. Although $1\%$ is too high an error rate for analysis of rare events on large datasets, it can be acceptable for other types of analysis. 
We were able to recover performances on a dataset composed of images from \cite{lugagne2019delta} (which have different aspect from DST) by a fine-tuning procedure with a training dataset as little as 3\% of the size of DST. This also shows that the chosen hyper-parameters are valid for an independent dataset.

\begin{table}
    \caption{Percentage of errors on dataset DSE2 (cells with different aspect from the training set).}
    \begin{tabular}{L{0.25}C{0.14}C{0.14}C{0.14}C{0.14}C{0.14}}
        \hline
        \multirow{2}{*}{\bfseries Method/Model}&
        {\bfseries Tracking}&
        \multicolumn{3}{c}{\bfseries Segmentation Errors}&
        \multirow{2}{*}{\bfseries Total}\\\cline{3-5}
        &{\bfseries Links}&{\bfseries Division}&{\bfseries False $-$}&{\bfseries False $+$}&\\
        \hline
        BACMMAN & $0.84$ & $0.59$ & $6.6$ & $0$ & $8.1$ \\
        DeLTA & $6.3$ & $1.1$ & $0.20$ & $0.022$ & $7.6$ \\
        Stacked Hourglass & $0.27$ & $0.64$ & $0.11$ & $0$ & $1.0$ \\
        Self-Attention & $0.30$ & $0.57$ & $0.068$ & $0$ & $0.93$ \\
        [0.5ex]
        \hline\\
    \end{tabular}
    \label{tableevaldiff}
\end{table}

To test whether our model was leveraging sequential information for segmentation or not, we trained a model with a similar architecture to the self-attention model that performed segmentation of a single frame through regression of the EDM. We found $3$ times more division errors on DSE1, which suggests that to segment bacteria at a given frame, our models are able to use information contained in the previous frame.  

\subsection{Ablation Experiments} 
\label{siablation}
To better understand the contribution of the main design choices of the self-attention network, we performed ablation experiments that are summarized in Table~\ref{sitableablation}. 

\begin{table}
    \caption{Ablation experiments. Percentage of errors on dataset DSE1 for different models. {\itshape No attention}: same model as the self-attention model with the attention layer replaced by a 3x3 convolution, which results in a model with the same number of parameters. {\itshape Half filters}: same model as the self-attention model with two times less initial filters (see~\ref{sitraining}). {\itshape No category}: self-attention network that only predicts the displacement and the EDM. The comparison with the self-attention model is indicated in parenthesis for the most different variables.}
    \begin{tabular}{L{0.25}C{0.15}C{0.14}C{0.14}C{0.16}C{0.13}}
        \hline
        \multirow{2}{*}{\bfseries Modification}&
        {\bfseries Tracking}&
        \multicolumn{3}{c}{\bfseries Segmentation Errors}&
        \multirow{2}{*}{\bfseries Total}\\\cline{3-5}
        &{\bfseries Links}&{\bfseries Division}&{\bfseries False $-$}&{\bfseries False $+$}&\\
        \hline
        No attention & \boldmath$0.10$ (x23) & $0.023$ & $0.0095$ & \boldmath$0.34$ (+0.34) & $0.47$ \\
        Half filters & \boldmath$0.011$ (x2.6) & $0.031$ (x1.6) & $0.0058$ & $0$ & $0.047$\\
        No category & $0.0064$ & $0.027$ (x1.4) & $0.0019$ & $0.0019$ & $0.037$\\
        \hline
    \end{tabular}
    \label{sitableablation}
\end{table}

Interestingly, when replacing the self-attention layer by a convolution layer (referred to as {\itshape No attention} in Table~\ref{sitableablation}), tracking performances drop and the increase of segmentation errors corresponds exclusively to false positives in empty microchannels. It suggests the self-attention layer is mainly useful for tracking, but also used to detect void microchannels certainly because the self-attention layer allows each area to take into account the whole image. 

Moreover, we trained a model with one more contraction/up-sampling level, and observed that the self-attention layer is more critical when there are less contractions/up-sampling. Therefore, including a self-attention layer allows both to improve performances and reduces the overall complexity of the model.

We also trained a network similar to the self-attention model with two times less initial filters, (referred to as \emph{Half filters} in Table~\ref{sitableablation}), and we observe twice as many errors as the base self-attention model.

Finally we analysed the contribution of the category prediction (Fig.~\ref{figunet}-(e-h)) by training a self-attention network that only predicts the displacement and the EDM (referred to as {\itshape No category} in Table~\ref{sitableablation}). We observe a small increase of division errors when categories are not predicted, which shows that category prediction is not an essential aspect of our method. 

\subsection{Attention mechanism} 
\label{siattentionmeca}
In order to get insights into the way the DNN uses the attention layer for tracking, we studied the attention weights. 

\begin{figure}[!ht]
    \centering
    \def\svgwidth{\textwidth}
    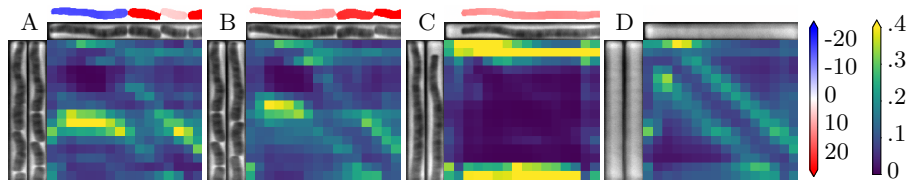
    \caption{Attention Weights. Each panel is composed of the attention weight matrix, on its left the predicted EDM for previous and current frame and at the top the predicted displacement and predicted EDM for current frame. After 4 contraction levels, the spatial dimensions of the input of the self-attention layer are 2x16, in order to focus on attention along the Y-axis, attention weights were summed along X-axis, resulting in a 16x16 weight matrix. Panel B corresponds to the next frame of panel A. Upper color bar: displacement along Y-axis; display range is [-25--25] pixels for panels A--D. Lower colorbar: matrix weights; display range is [0--0.4]}
    \label{sifigattentionweights}
\end{figure}

A weight matrix can be read in the following way: for each output region (columns), it shows where the attention was mostly focused on the inputs (rows). For instance, a perfect diagonal attention matrix would mean that most of the information needed to produce an output region comes from the same input region.

Weight matrices for different frames of DSE1 are shown in Fig.~\ref{sifigattentionweights}.
We observe that: 
\begin{itemize}
    \item Weights are usually lower at the diagonal than at its direct surroundings, this is likely a consequence of the skip connection.
    \item There is a bias towards higher weights above the diagonal which reflects the mostly downward movement of bacteria due to growth.
    \item The self-attention layer effectively allows to integrate global context by allowing each area of the image to take into account the rest of the image.
    \item In the case of long cells, self-attention focuses on the edges of cells rather than on their interior (See Fig.~\ref{sifigattentionweights}-C).
    This is particularly visible, in Fig.~\ref{sifigattentionweights}-A where a division occurs, and the attention for upper daughter cell prediction is focused on the edges of the mother cell. Fig.~\ref{sifigattentionweights}-B corresponds to the next observation, and consistently, we observe that the attention moves upwards and is still focused on the edges of the previous cell. 
\end{itemize}

\section{Conclusion}

In this study we present a novel formulation of the multi-object tracking problem using DNN. In contrast with most existing methods, tracking is done through a single DNN step, allowing significant speed improvement. Moreover it performs jointly segmentation and tracking, allowing to leverage sequential information for segmentation. To support this formulation, we introduced a DNN architecture based on U-Net modified by incorporating a self-attention layer.

We applied successfully this method to the problem of bacteria growing in the {\itshape mother machine}, and achieved error rates lower than $0.005\%$ for tracking and of $0.03\%$ for segmentation, outperforming current state-of-the-art methods, and making this method well-suited for high-throughput data analysis. 

The simplicity of our formulation and our model allows to adapt it easily to other problems with other geometry or other objects to be detected, and to implement it using other DNN architectures.

\subsubsection{Availability}\setcounter{footnote}{0}
DistNet is publicly available: we provide source code\footnote{\url{https://github.com/jeanollion/distnet}} as well as a module for Bacmman software\footnote{\url{https://github.com/jeanollion/bacmman/wiki/DistNet}} and a tutorial for fine-tuning that includes a ready-to-use notebook\footnote{\url{https://github.com/jeanollion/bacmman/wiki/FineTune-DistNet}}.

\subsubsection{Acknowledgements}
We thank Lydia Robert for kindly providing us with the datasets, as well as Thomas Robert and Hervé Mari-Nelly for the useful comments on the manuscripts. We would also like to thank the reviewers, in particular reviewer 2 for the constructive critical comments that helped us improve the manuscript.

\bibliographystyle{splncs04} 
\bibliography{refs}

\renewcommand{\theequation}{S\arabic{equation}}
\renewcommand{\thefigure}{S\arabic{figure}}
\renewcommand{\thesection}{S\arabic{section}}
\renewcommand{\thetable}{S\arabic{table}}

\pagebreak
\title{Supplementary Materials for DistNet: Deep Tracking by displacement regression: application to bacteria growing in the {\itshape Mother Machine}}

\titlerunning{SM for DistNet: Deep Tracking by displacement regression}

\author{Jean Ollion\inst{1}\inst{2}\orcidID{0000-0002-9182-4624} \and Charles Ollion\inst{3}\orcidID{0000-0002-6763-701X}}
\authorrunning{J. Ollion and C. Ollion}
%
\institute{
Laboratoire Jean-Perrin UMR 8237, Sorbonne Universit\'e, Paris, France 
\email{jean.ollion@polytechnique.org} 
\and 
SABILAb, Paris, France
\and
Heuritech, Paris, France \\ \email{charles.ollion@gmail.com}}
\maketitle              
\section{Self-Attention layer} \label{sisa}
The self-attention layer can be seen as a set-to-set function $\mathbf{f}$ which applies to each $h_i$ and incorporates the context of each element $\{ h_i \}$ through self-attention:
$$h_i^{out} = \mathbf{f}(h_i, \{ h_i \})$$
Each element is first transformed through three learnt dense projection layers to Q (query), K (key), and V (value) vectors of dimension $d_k$ in the following way: 
$$Q_i, K_i, V_i = \mathbf{W^q}h_i+b^q, \mathbf{W^k}h_i+b^k, \mathbf{W^v}h_i+b^v$$
These vectors enable to compute the self-attention in the following way:
$$\large{Attention(Q, K, V) = softmax_k(\frac{QK^T}{\sqrt{d_k}}) V} $$
Finally, to compute the output of the layer, we apply an affine output transformation to the output attention:
$$h_i^{out} =  \mathbf{W^o} \sum_k Attention(Q_i, K_k, V_k) + b^i$$
This output is concatenated with the initial input vectors followed by a 1x1 convolution layer.

Intuitively, the set of elements $ \{ h_i \}$ is transformed into a new set $ \{ h_i^{out} \}$ which takes into account all other elements, effectively mixing information. However, the relative order of elements is irrelevant here, as they are treated as a {\itshape set}. In order to include the relative position of elements, we add a special positional embedding to the input $h_i$, by projecting the index $i$ to a vector of similar dimension as $h_i$: 
$$h_i^\prime = h_i + Embedding(i)$$

\begin{figure}[!ht]
    \centering
    \def\svgwidth{0.8\textwidth}
\begingroup%
  \makeatletter%
  \providecommand\color[2][]{%
    \errmessage{(Inkscape) Color is used for the text in Inkscape, but the package 'color.sty' is not loaded}%
    \renewcommand\color[2][]{}%
  }%
  \providecommand\transparent[1]{%
    \errmessage{(Inkscape) Transparency is used (non-zero) for the text in Inkscape, but the package 'transparent.sty' is not loaded}%
    \renewcommand\transparent[1]{}%
  }%
  \providecommand\rotatebox[2]{#2}%
  \newcommand*\fsize{\dimexpr\f@size pt\relax}%
  \newcommand*\lineheight[1]{\fontsize{\fsize}{#1\fsize}\selectfont}%
  \ifx\svgwidth\undefined%
    \setlength{\unitlength}{467.9523495bp}%
    \ifx\svgscale\undefined%
      \relax%
    \else%
      \setlength{\unitlength}{\unitlength * \real{\svgscale}}%
    \fi%
  \else%
    \setlength{\unitlength}{\svgwidth}%
  \fi%
  \global\let\svgwidth\undefined%
  \global\let\svgscale\undefined%
  \makeatother%
  \begin{picture}(1,0.49439039)%
    \lineheight{1}%
    \setlength\tabcolsep{0pt}%
    \put(0,0){\includegraphics[width=\unitlength,page=1]{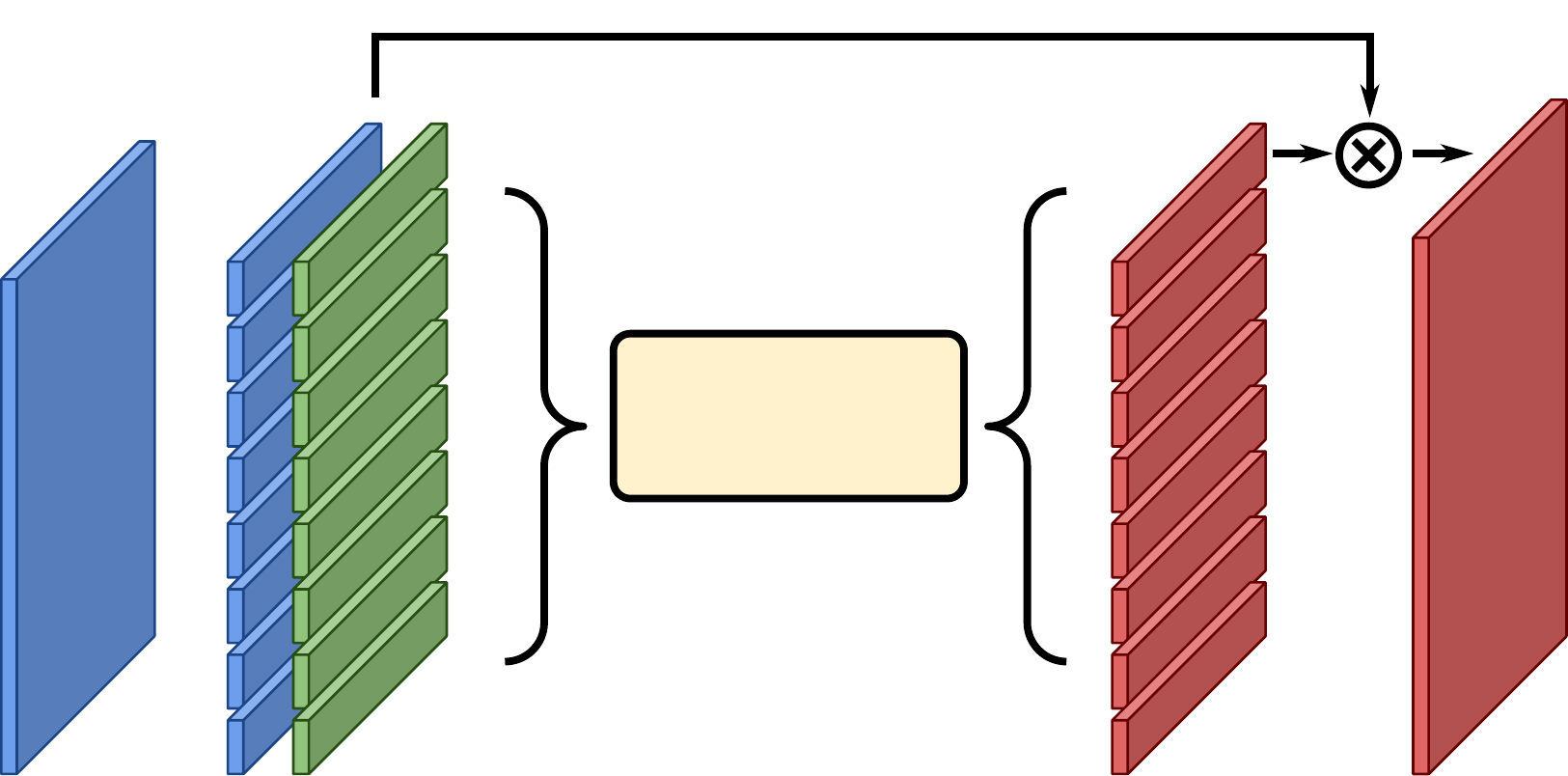}}%
    \put(0.50656334,0.23875882){\color[rgb]{0,0,0}\makebox(0,0)[t]{\lineheight{1.25}\smash{\begin{tabular}[t]{c}Self-Attention \\Layer\end{tabular}}}}%
  \end{picture}%
\endgroup%

    \caption{Self-attention layer. A spatial feature map (left, blue) is interpreted as a set of feature vectors. These vectors are combined with positional embedding that only depend on their index (for instance $i \in [0,7]$ if the spatial dimensions are $8 \times 1$). The self-attention effectively transforms the set into a new one, where global information may be used. The final output has a skip connection with the input and can be re-interpreted as a spatial map.}
    \label{sifigsa}
\end{figure}

\section{Dataset acquisition}
Images from the study \cite{robert2018mutation} were used for datasets DST and DSE1. Dataset DSE2 is composed of publicly available datasets published along with software for {\itshape mother machine} analysis \cite{sachs2016image,lugagne2019delta,kaiser2018monitoring}.

\section{Stacked hourglass network architecture} \label{sistacked}
Another way to integrate the information at global scale could be stack multiple U-Nets as the intermediate representations could mix information with more and more context, up to the point where we could consider this context large enough to represent what happens at global scale within the image. This technique has already been used in the context of human pose estimation and is usually referred as Stacked Hourglass Networks (the term {\itshape hourglass network} refers to an encoder-decoder network structure such as as U-Net) \cite{newell2016stacked} in which the decisions to output a human skeleton keypoint must be very dependant on the global pose of the human, and cannot be taken independently within a small pixel context. Stacking the hourglass networks effectively capture various spatial relationships and dependencies within large objects (a body in the case of Human pose estimation, or bacteria in the case of {\itshape mother machine} data).

\begin{figure}[!ht]
    \centering
    \def\svgwidth{\textwidth}
    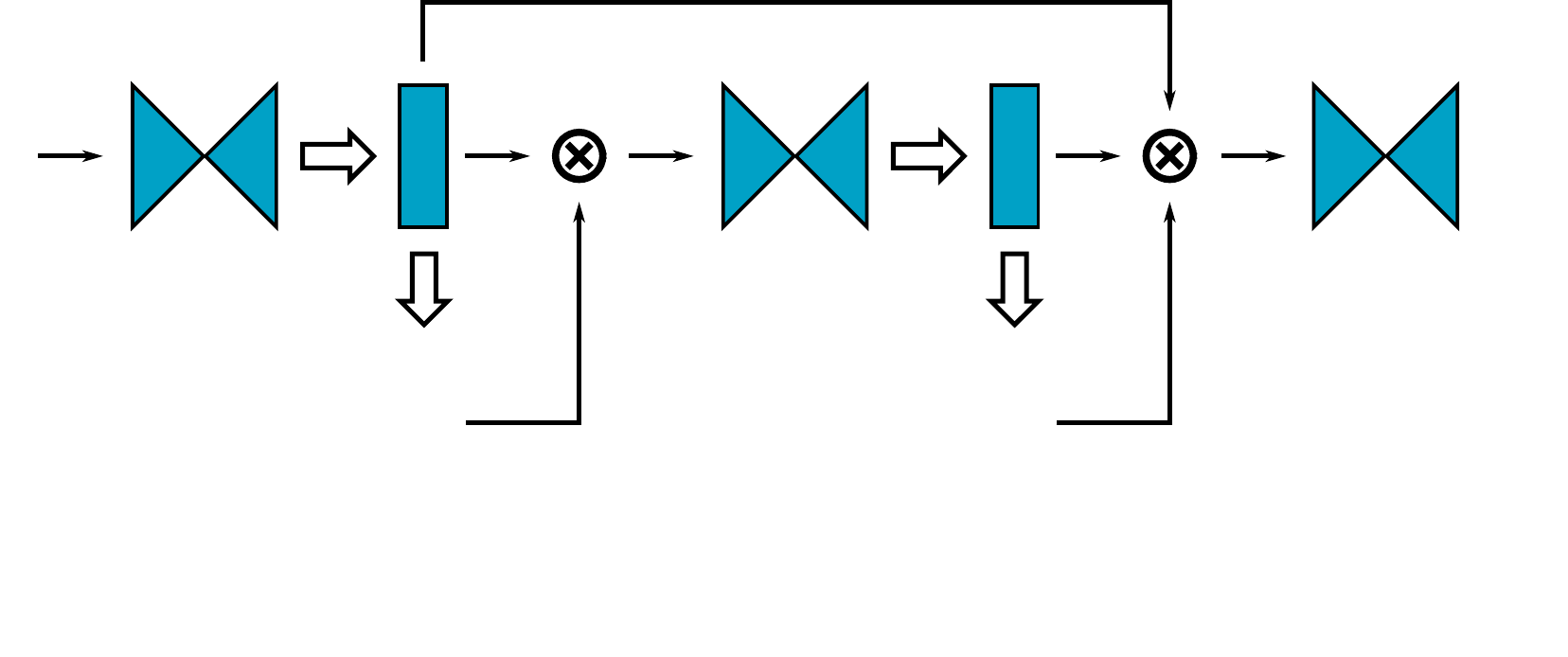
    \caption{Stacked hourglass network architecture. Each hourglass module produces multi-channel feature maps, from which the output are computed. Then losses are applied on those outputs. They are concatenated with the feature maps and the feature maps from the previous module to feed the next hourglass module.}
    \label{sifigstacked}
\end{figure}

The network is essentially similar to stacked U-Nets modules (as in Fig.~\ref{figunet}). A projection to the actual final layers of the task is performed at each intermediate step, and the loss is computed both at intermediate steps and final output of the network, which is often referred to as intermediate supervision. The input of a given module is the concatenation of the output feature maps of the two previous modules, as well as the intermediate output of the previous module (See Fig.~\ref{sifigstacked}).

\section{Training} 
\label{sitraining}
The architectures of our models are based on the original U-Net architecture with some slight changes: in order to limit the number of parameters, we set a limit to the channel number of feature maps. We also replaced all the 3$\times$3 convolution layers following concatenation layers at each up-sampling step by 1$\times$1 convolutions. The last 3$\times$3 convolution layer of the decoder is followed by a 1$\times$1 convolution layer with the same number of channels.
The Stacked hourglass model is composed of 4 stacked U-Net modules of that have 64 channels at the first level and a maximum of 512 channels, 5 contractions/up-sampling levels with a total of 72.6M parameters.
The Self-attention model has 128 channels at the first level and a maximum of 1024 channels, 4 contractions/up-sampling levels, with a total of 52.6M parameters. The last contraction of the encoder is followed by a 3$\times$3 convolution layer, a self-attention layer and a 20\% dropout layer.

Models were implemented with keras/tensorflow, trained using Adam optimizer with a learning rate of $2.10^{-4}$ and reduced on plateau of 5 epochs by a factor 2 until $10^{-6}$. Total number of epoch was about 100 and batch size was set to 64.

For EDM regression, L2 loss was used, for displacement regression, L1 loss was used and for categories a weighted sparse categorical cross entropy loss was used with a weight per category to compensate the lower frequency of the two classes dividing cell and cell not linked to a previous cell. 

\section{Data augmentation} 
\label{sidataaug}
Data augmentation used in this work can can be grouped into two categories: illumination transformations and geometrical transformations. 

\subsubsection{Illumination transformations} 
In order to simulate variations in signal to noise ratio, we added a mixture of 3 models classically used to model noise in microscopy images: Gaussian additive noise, Gaussian multiplicative noise and Poisson noise each one with a random level. We also performed an elastic deformation of the histogram and added a random intensity variation along the Y-axis in order to simulate variations in illumination among different imaging setups as in \cite{lugagne2019delta}. 
Last, we set a random minimal and maximal intensity value in the range $[0, 1]$ with a minimal range of $0.1$. At prediction, input images are normalized in the range $[0, 1]$. Setting a random intensity range within $[0, 1]$ during training aims at increasing robustness to the presence of long tails in intensity distribution, which is very common in phase-contrast images.
All theses transformations were performed with the same parameters on each couple of successive input images. 

\subsubsection{Geometrical transformations}
We used standard affine transformations: shifting, scaling, shear transform, rotation and horizontal flipping. Transformation parameters were randomly chosen and we added constraints to ensure the combination of transformations was not producing unrealistic images: we limited the ratio of scaling between X and Y axis so that bacteria were not too thin or not too round, and we limited the combination of horizontal shift, rotation and scaling so that bacteria located at the ends of the microchannels did not go out of the image.
The bacteria strands we used in our training dataset were not able to swim within the microchannels thus their only displacement was due to growth. In order to enable the network to track bacteria that are able to swim, we simulated swimming by translating the image between two successive randomly chosen bacteria, by a random distance in the direction of the open-end.
Cells going out of the image with an observed length inferior to 20 pixels where erased. 
Only scaling and shearing were performed with the same parameters on each couple of successive input images.

\end{document}